\documentclass[10pt,letterpaper,conference]{IEEEtran}
\usepackage{amsmath,amssymb,amsfonts}
\usepackage{algorithmic}
\usepackage{textcomp}
\usepackage{xspace}
\usepackage{balance}
\usepackage{graphicx}
\usepackage[hyphens]{url}
\usepackage{epstopdf}
\usepackage{microtype}
\usepackage{booktabs}
\usepackage[table,xcdraw]{xcolor}
\usepackage{hyperref}
\usepackage[numbers]{natbib}
\usepackage{amsmath}
\usepackage{enumitem}
\usepackage{amssymb}
\usepackage{pifont}

\usepackage{balance}
\iffalse 
	
\else
	
\fi

\usepackage{multirow}
\newcommand{\PreserveBackslash}[1]{\let\temp=\\#1\let\\=\temp}
\newcolumntype{M}[1]{>{\PreserveBackslash\centering}m{#1}}
\newcolumntype{C}[1]{>{\PreserveBackslash\centering}p{#1}}
\newcolumntype{R}[1]{>{\PreserveBackslash\raggedleft}p{#1}}
\newcolumntype{L}[1]{>{\PreserveBackslash\raggedright}p{#1}}

\newcommand{\fakepar}[1]{\smallbreak\noindent}
\newcommand{\boldpar}[1]{\smallbreak\noindent\textbf{#1.}}

\newcommand{\fifteenfour}{\mbox{IEEE~802.15.4}\xspace}
\newcommand{\eleven}{\mbox{IEEE~802.11}\xspace}
\newcommand{\wifi}{\mbox{Wi-Fi}\xspace}

\usepackage{manyfoot}
\DeclareNewFootnote{A}
\DeclareNewFootnote{B}

\let\footnoteR\footnoteB
\let\footnote\footnoteA
\iftrue
    \newcommand{\mike}[1]{\footnoteR{{\color{red}\bf MB: #1}\color{red}}}
    \newcommand{\martin}[1]{\footnoteR{{\color{red}\bf MA: #1}\color{red}}}
    \newcommand{\jp}[1]{\footnoteR{{\color{red}\bf JP: #1}\color{red}}}
    \newcommand{\willian}[1]{\footnoteR{{\color{red}\bf WL: #1}\color{red}}}
    \newcommand{\anshul}[1]{\footnoteR{{\color{red}\bf AP: #1}\color{red}}}
    \newcommand{\francis}[1]{\footnoteR{{\color{red}\bf FB: #1}\color{red}}}
\else
    \newcommand{\mike}[1]{}
    \newcommand{\martin}[1]{}
    \newcommand{\jp}[1]{}
    \newcommand{\willian}[1]{}
    \newcommand{\anshul}[1]{}
    \newcommand{\francis}[1]{}
\fi

\begin{document}

\title{Towards Secure Wireless Mesh Networks\\ for UAV Swarm Connectivity:\\ Current Threats, Research, and Opportunities}
\author{\IEEEauthorblockN{Martin Andreoni Lopez,
Michael Baddeley, Willian T. Lunardi, Anshul Pandey and Jean-Pierre Giacalone }
    \IEEEauthorblockA{Secure Systems Research Center (SSRC),
    Technology Innovation Institute (TII)\\
    Abu Dhabi, United Arab Emirates\\
    \{martin, michael, willian, anshul, jean-pierre\}@ssrc.tii.ae}}
\maketitle

\begin{abstract}

UAVs are increasingly appearing in swarms or formations to leverage cooperative behavior, forming flying ad hoc networks. These UAV-enabled networks can meet several complex mission requirements and are seen as a potential enabler for many of the emerging use-cases in %6G, and 
future communication networks. Such networks, however, are characterized by a highly dynamic and mobile environment with no guarantee of a central network infrastructure which can cause both connectivity and security issues. While wireless mesh networks are envisioned as a solution for such scenarios, these networks come with their own challenges and security vulnerabilities. In this paper, we analyze the key security and resilience issues resulting from the application of wireless mesh networks within UAV swarms. Specifically, we highlight the main challenges of applying current mesh technologies within the domain of UAV swarms and expose existing vulnerabilities across the communication stack. Based on this analysis, we present a security-focused architecture for UAV mesh communications. Finally, from the identification of these vulnerabilities, we discuss research opportunities posed by the unique challenges of UAV swarm connectivity.
% Moreover, from the identification of these vulnerabilities, we discuss research opportunities posed by the unique challenges of UAV swarm connectivity and present a new security-focused architecture for UAV mesh communications. Finally, we present a new security-focused architecture for UAV mesh communications and open-source this to the research community.
\end{abstract}


\makeatletter
\def\ps@IEEEtitlepagestyle{%
    \def\@oddfoot{\mycopyrightnotice}%
    \def\@evenfoot{}%
}
\def\mycopyrightnotice{%
    \begin{minipage}{\textwidth}
        \centering\tiny{\copyright{} 2021 IEEE.  Personal use of this material is permitted.  Permission from IEEE must be obtained for all other uses, in any current or future media, including reprinting/republishing this material for advertising or promotional purposes, creating new collective works, for resale or redistribution to servers or lists, or reuse of any copyrighted component of this work in other works.}
    \end{minipage}
    \gdef\mycopyrightnotice{}% just in case
}

%\vspace{-2.0mm}
\section{Introduction}
\label{sec:introduction}
%I suggest to read this small piece of beauty from James Kurose
%http://www-net.cs.umass.edu/kurose/writing/intro-style.html

The precise location of objects, real-time mapping of unknown or difficult environments, and task time reduction when compared with single devices make the UAV swam attractive for commercial and military use. In addition, UAV swarms can help to facilitate the requirements of future networks, providing high availability, greater connection density and low end-to-end delay in the form of Aerial Access Networks~\cite{dao2021survey}. Such scenarios, however, introduce significant challenges in the form of device mobility, ad hoc network formation, and limited device energy, that should be combined with self-organization, robustness through redundancy, and task coordination requirements of UAV Swarms.

%UAV swarms can help to facilitate the requirements of future networks, providing high availability, greater connection density and low end-to-end delay in the form of Aerial Access Networks~\cite{dao2021survey}. Such scenarios, however, introduce significant challenges in the form of device mobility, ad hoc network formation, and limited device energy. 

Existing wireless mesh technologies have the potential to address these issues, and both the \eleven~\cite{80211_ieee} and \fifteenfour~\cite{802154_ieee} standards have popularized mesh protocols for household and industrial use-cases. Indeed, while \fifteenfour is a wide-ranging standard typically focusing on low-power, low-rate communications, the IEEE\,802.11s and Independent Basic Service Set (IBSS) amendments included as part of the \eleven-2012 revision enable \wifi-based mesh networks capable of supporting high throughput applications across the mesh, such as image or video -- a common focus of many UAV scenarios.

%Nevertheless, despite considerable advances in performance, scalability, and hardening of security,
Nevertheless, despite considerable advances mesh communications still face several of the complex challenges identified within seminal works~\cite{gupta2000capacity, akyildiz2005wireless}. Specifically, within the context of this paper, we identify key weaknesses in current state-of-the-art (SOTA) mesh solutions that present critical security challenges within the context of UAV swarms -- a relatively new research area in comparison to much of the initial mesh research. For example, while distributed routing protocols can adapt relatively easily to static or slow topology changes in community or industrial mesh networks, the highly mobile environments of UAV swarms, with speeds ranging from 30 to more than 400 km/h~\cite{arafat2019routing}, present a significant challenge. 

We argue that the unique conditions arising from UAV swarm connectivity requires a comprehensive analysis of security and resilience in wireless mesh networks, with key considerations for future research. %}
In this paper, we aim to survey the current SOTA of secure UAV Swarm communications and examine how underlying assumptions in current wireless mesh protocols and standards invite key vulnerabilities and expose significant threats to UAV swarm communications. We also provide a comprehensive overview of these security challenges, from the physical to the application layer, showing how current solutions leave the entire communication system subject to different attack vectors. While we describe specific threat areas that can be mitigated through existing techniques, there is significant exposure within Layers 1 (physical) to 3 (networking) that remains unsolved with current solutions. We identify key research challenges and opportunities and propose a security-focused mesh networking architecture for secure UAV mesh communications that can be used by the research community to address vulnerabilities in UAV swarm communication.

This paper is structured as follows. In Section~\ref{sec:related_work} we introduce background on wireless mesh communications and recent research on routing protocols for UAV mesh networks. Section~\ref{sec:threats} presents the main threats and vulnerabilities on UAV swarm communication, from the physical to the application layer. An architecture that addresses the mentioned issues is presented in Section~\ref{sec:newarch}. 
In Section~\ref{sec:challenges} we discuss the current and future research, open issues, and challenges. Finally, Section~\ref{sec:conclusions} concludes our work.

% \vspace{-2.00mm}
\section{Background}
%\textcolor{blue}{(Michael, Willian, ...)}}
\label{sec:related_work}
% Brief examination of Wireless Mesh Networks and their issues.
Fundamental capacity limitations complicate the application of mesh networks within UAV use-cases. As highlighted in~\cite{gupta2000capacity}, when considering half-duplex radios (such as those commonly found in commercial drones) every forwarding action by an intermediate node reduces the maximum end-to-end wireless throughput by at least $1/N$. Furthermore, routing nodes serving as relays for nodes further away in the mesh are forced to share their bandwidth with child branches. Such bottlenecks present difficulties for the network in responding to changing topology or react to threats. Therefore, while protocols have evolved to mitigate these challenges, the distribution of intelligence within the mesh is a complex trade-off between optimization and scalability. Centralizing control allows optimal solutions with global visibility but generate considerable overhead when forwarding control signaling across multiple hops. Distributing control solves this issue and improves responsiveness, but then suffers from reduced visibility of the network state~\cite{baddeley2020thesis}. 

The authors of~\cite{siddiqui2007security} present high-level overviews on security issues in wireless mesh networks: identifying how  mesh constraints such as processor performance, battery life, mobility, and limited bandwidth present unique security challenges. Many of these constraints occur due to the complexities of routing information across a distributed multi-hop network.  However, while many routing protocols have been proposed for wireless mesh networks, most approaches neglect to address security.% aspects.

Better Approach To Mobile Ad-hoc Networking (BATMAN)~\cite{ietf_batman},
and Optimized Link State Routing (OLSR)~\cite{olsr} 
% MB: I took out the acronyms for space as we are just citing works here. Original text below
are two widely used routing protocols that assume security will be implemented in the upper layers~\cite{arafat2019routing},
% Moreover, protocols such as ARAN~\cite{sanzgiri2005authenticated}, SAODV~\cite{zapata2002secure}, and Secure OLSR use
while asymmetrical key encryption techniques sign every message, but consume resources that can be expensive in UAV swarm networks~\cite{sahingoz2013multi}. 
% Some solutions use symmetric keys such as secure on-demand routing protocol Ariadne~\cite{hu2005ariadne}, SEAD~\cite{hu2002secure} and 
While symmetric key solutions such as CASTOR~\cite{galuba2010castor} avoid the high processing of asymmetrical encryption, they require secure key exchange between source and destination, which can be complex to implement in a highly dynamic UAV swarm environment. PASER is a position-aware routing protocol specifically focused on UAV wireless mesh networks~\cite{sbeiti2015paser}. PASER uses hybrid cryptography schemes, asymmetric key for mutual node authentication and key exchange, and symmetric for routing message authentication solving inherent mesh wireless network issues such as blackhole and wormhole attacks. Unfortunately, protocols using Public Key Infrastructure (PKI) assume a Certification Authority (CA) to provide certificates with the public and the private key. While this assumption is feasible within the UAV swarm (since the CA can be accessed through connection to the cloud), it is desirable to avoid a centralized solution to support a fully autonomous swarm and mitigate loss in connectivity to a backbone infrastructure.  JarmRout is a jamming-resilient multipath routing protocol, based on three features: i)~link quality scheme, analyzing received signal strength indication (RSSI) of received packets; ii)~traffic load scheme, the packets stored in the queues;  and iii)~spatial distance between source and destination~\cite{pu2018jamming}.%MA: Adding from here
This work only focus on secure routing without providing any solution to other secure aspects. 

While the flexibility %afforded 
offered by wireless mesh makes such networks an ideal candidate for intra-drone communication in UAV swarms, these fundamental challenges present vulnerabilities within SOTA mesh networks that introduce many attack
vectors that bad actors could exploit.

\begin{table*}[htbp]
	\caption{Summary of Threats and Vulnerabilities for UAV Mesh Swarm Connectivity.}
	\label{table:threats_and_vulnerabilities}
	\renewcommand{\arraystretch}{1.1}
	\centering
    \begin{tabular}{ c L{3cm} L{5cm} L{5cm} }
    \toprule
      	\bfseries Layer & \bfseries Threat/Vulnerability & \bfseries Exploits & \bfseries Defense/Opportunity \\ 
      	\midrule
        \multirow{2}{*}{Physical} 
  	        & Passive Eavesdropping & Broadcast Nature of Wireless Channels  & Strong Encryption \\
  	        & RF Jamming & Wireless contention & Avoid Interference Zones \\
  	    \hline
  	    \multirow{2}{*}{Link} 
  	        & Spoofing & Intentional Collision & Intrusion Detection Systems (IDS) \\ 
  	        & Frame Modification & MAC Spoofing & Encryption \\
  	    \hline
  	    \multirow{2}{*}{Network} 
  	        & Routing Forwarding & Selfish Attacks & Intrusion Detection Systems (IDS) \\
  	        & Data Forwarding & Collusion Attacks& Firewall \\
  	    \hline
  	    \multirow{2}{*}{Transport} 
  	        & Packet Corruption & Denial of Service (DoS) & Intrusion Detection Systems (IDS) \\
  	        & Protocol Weakness & Session Hijacking & Transport Layer Security (TLS) \\
  	    \hline
  	    \multirow{2}{*}{Application} 
  	        & ROS2 Bugs & Malware & Authentication \\
  	        & Open Protocols & Injection/Modification & Encryption \\
    \bottomrule
    \end{tabular}
    \vspace{-3.00mm}
\end{table*}

\begin{figure}
    \centering
    \includegraphics[width=1.\columnwidth]{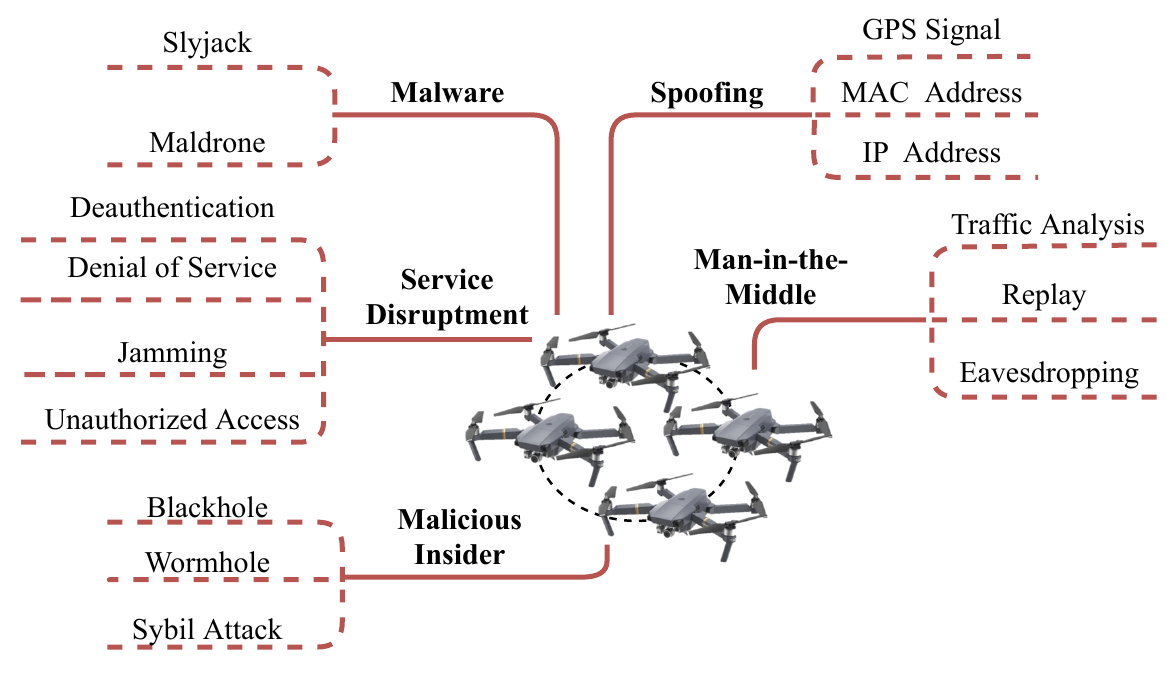}
    \caption{Classification of threats faced by UAV wireless mesh networks.}
    \vspace{-4.00mm}
    \label{fig:threats_classification}
\end{figure}

%Anshul:  Will be adding the possible attacks at the Physical layer apart from jamming. The test will be in blue color fo easy in identification. We can dicuss later and come to a common draft.
% \section{Vulnerabilities And Threats\textcolor{blue}{(Martin, Michael, Anshul...)}}
\section{Vulnerabilities And Threats}
% \textcolor{blue}{(Please read it and add comments if necessary)}} 
\label{sec:threats}
% MB: Too long, rewording
% Different protocols are used on the UAV mesh communication stack from the physical to the application layer. Most of these protocols, however, are designed with no security or they assume that security issues are handled on the above layers. This assumption, and the absence of a central controller on autonomous UAV swarms, make the protocols at the \textit{physical}, \textit{link}, \textit{network} and \textit{transport} layers vulnerable to several threats. The attacker's objective is to disrupt the communication between drones in the swarm. Thus, without access to the network, an external attacker can control or disrupt communications within the network -- having more power and a wider communication range than benign nodes. On the other hand, the attack can be internal, once the attacker compromises benign nodes or network credentials through cryptanalysis or physical attack. Within this section, we systematically identify the vulnerabilities faced by UAV mesh communications at each networking layer and show how exploitation of these weaknesses presents a specific threat. We classify these in Figure~\ref{fig:threats_classification}, and provide a high-level summary in Table~\ref{table:threats_and_vulnerabilities}.
Many mesh protocols are designed with no security or assume security issues are handled on the above layers. This assumption, and the absence of a central controller, make the protocol layers vulnerable to several threats. Suppose an attacker's objective is to disrupt the communication between drones in the swarm. Thus, without access to the network, an external attacker can control or disrupt communications within the network. On the other hand, an attack can be internal -- once an attacker compromises benign nodes or network credentials through cryptanalysis or physical attack. We systematically identify vulnerabilities faced by UAV mesh communications at each networking layer and show how exploitation of these weaknesses presents a specific threat. We classify these in Figure~\ref{fig:threats_classification}, and provide a high-level summary in Table~\ref{table:threats_and_vulnerabilities}.

% The authors of~\cite{yaacoub2020security} and~\cite{nassi2019sok} examine the security challenges presented by the use of drones in civilian and military contexts, as well as the anti-drone methods available to mitigate against malicious drone attacks -- all of which are covered by the threat vectors identified within this paper.

\subsection{Physical Layer}
\label{sec:threats_physical}

\boldpar{Vulnerabilities}
The broadcast nature of the wireless channel makes transmissions at the physical layer susceptible to various attacks and malicious activities. Broadly, the threats at the physical layer can be classified as either passive or active attacks~\cite{ada_passive}. Passive attacks are mainly categorized as eavesdropping and traffic analysis, while active attacks mainly include jamming and interference. There may also be a case of hybrid attack where the malicious node may behave according to the wireless channel conditions. All the attacks presented at the physical layer are considered external to the UAV swarm.

\boldpar{Threats} 
In a multi-hop UAV swarm communication, eavesdropping is the simplest form of passive attack to perform.  The attacker only needs to overhear the information broadcast over the wireless channel. This may expose confidential network information. %Like eavesdropping, 
Traffic analysis is another type of passive attack at the physical layer. Herein, the attacker is not interested in the data. Instead, the attacker looks to capture network information such as the location of key network entities, routing, and other involved behavior patterns to carry out more sophisticated attacks. The challenge is to detect such passive attacks, which are generally mitigated through strong encryption.

Active attacks can exploit the physical layer by intentionally transmitting interfering signals. In particular, Radio Frequency (RF) jamming attempts to occupy legitimate channels and degrade the received signal-to-interference noise ratio (SINR). In the context of UAV mesh networks, such disruption could lead to loss of geospatial positioning, mission references, and even flight control~\cite{adaptive_jam}. While many wireless protocols are built to withstand external interference from other devices and networks, those widely used in UAV swarm communication typically %(though not exclusively)
operate over the 2.4\,GHz ISM band, where deliberate and targeted jamming attacks can quickly disrupt communications. For the purposes of this paper we broadly categorize jammers into three types of attack: \textit{constant}, \textit{reactive}, and \textit{cognitive},% however 
for more in-depth analysis of RF jammers refer to~\cite{sufyan2013detection}. 
% Table~\ref{table:jammer_types} summarizes the main features of the RF jamming attacks. 

\textit{Constant} jammers are low-cost, and generate a constant or sweeping tone across a $W\,$MHz channel, allowing the jammer full coverage of a spectrum band within a short period of time. Communication schemes that employ frequency diversity through Frequency Hopping Spread Spectrum (FHSS), however, provide some defense. Within a UAV swarm, this (typically) low-powered attack can also be mitigated through spatial diversity. 

\textit{Reactive} jammers sense the spectrum before transmitting a blocking signal in response to a transmission from a nearby device. Thus, the jamming device can conserve energy -- as a high-powered radio transmission will typically use more power than passively receiving. 

\textit{Cognitive} jammers are typically high-powered, military-grade jamming solutions able to actively monitor the spectrum for active communications and try to disrupt the underlying communications: such as beating the Forward Error Correction (FEC) schemes at the PHY and MAC level. Such sophisticated attacks are extremely difficult to detect as opposed to a constant tone. For instance, the attacker may listen passively when the wireless channel condition is favorable. If the channel is poor, it may disrupt the network communication by introducing the interfering signals~\cite{adaptive_phy}.

\subsection{Link Layer}
\label{sec:threats_link}
%\vspace{-2.00mm}

\boldpar{Vulnerabilities}
The link layer is responsible for multiplexing data streams, error control, and medium access control. The main vulnerability at this layer lies in MAC Address Spoofing, where one station claims the identity of another member of the network. Other frame-level exploits and vulnerabilities include sniffing and broadcast storms. 

\boldpar{Threats}
\textit{Link-layer jamming attacks} are a form of Denial of Service (DoS). The attacker repeatedly transmits the MAC frame header without payload. The benign node subsequently always finds the channel busy and will back-off for some time before sensing again. A more sophisticated version can be used to monitor the network activity of other layers and mimic the typical behavior such as packet size, protocol patterns, etc.

An \textit{Intentional Collision Attack} is a feature of the link-layer protocol implementation. When two nodes attempt to transmit at the same frequency simultaneously, a packet collision occurs. In collision-avoidance MAC protocols (such as in \eleven), an exponential back-off is applied at both nodes. The attacker can take advantage of this feature and send continuously corrupted MAC packets. Both these attacks are considered external to the UAV swarm.

A \textit{MAC Spoofing Attack} is a common wired and wireless link layer attack. As a MAC address is a unique identifier provided by the manufacturer, it is often used as an authentication factor to grant network privileges to a user. Wireless standards such as \eleven do not provide any security against a MAC source address modification in the transmitted frames. %Modifying the MAC source address in the frame is considered spoofing.
MAC spoofing is a way to evade Intrusion Detection Systems (IDS) and Access Control lists, enabling the attacker to masquerade as a legitimate user and gain access to the network. 

\textit{Replay Attacks}, also known as Man-in-the-Middle (MitM) can be executed for internal or external nodes. Normally, this attack is executed together with MAC spoofing and eavesdropping. Initially, a node C will eavesdrop on the broadcast communication between two nodes, A and B. Node C will then spoof the MAC address to impersonate node B when communicating with node A.

%------------------------------------------------------------%
%http://kresttechnology.com/krest-academic-projects/krest-mtech-projects/CSE/M.%20Tech%20Basepapers%202017-18/network%20n%20security/13.%20PASER%20Secure%20and%20Efficient%20Routing%20Approach.pdf
\subsection{Network Layer}
\label{sec:threats_network}

\boldpar{Vulnerabilities}
% The network layer is responsible for the data transmission through the transport layer. 
This layer is responsible for IP packet encapsulation and routing of information across the network.
Specifically, in wireless mesh networks, by using multi-hop routing algorithms. Attacks to this layer target the routing mechanism of the UAV swarm communication and affect the packet forwarding~\cite{maxa2017survey}.

\boldpar{Threats} 
A \textit{Black-hole attack} consists of corrupting route discovery in the routing protocol to place the malicious node on the path of as many routes as possible. Once done, the attacker node may drop all or the majority (gray-hole attack) of data packets and thus significantly degrade routing performances. 

A \textit{wormhole attack} consists of two colluding nodes creating a wormhole, or a tunnel, within the network by corrupting the route discovery process to capture sensitive data. Both techniques imply forging false routing packets. 

A \textit{Sybil Attack} is named after the subject of the book Sybil, a case study of a woman diagnosed with multiple fake identities. These fake identities are known as Sybil nodes. The Sybil nodes can outvote the honest nodes in the system. Usually, %peer-to- peer 
multi-path systems are vulnerable to Sybil attacks. Cryptography is a broadly used technique to secure the underlying routing protocols against nodes’ misbehavior by ensuring the integrity and authenticity of routing packets.  Nevertheless, cryptography consumes significant network resources like processing, delay, and throughput and it can affect energy consumption in UAV networks~\cite{energy-encryption}. All these attacks need to have at least one insider node to perpetrate the attack. 

Nevertheless, the \textit{Hello Flood Attack} can be performed for an external node. Nodes are required to send messages to other nodes to show proximity, routes, performance, among other metrics to decide the best path for communication. An attacker can flood the channel with small \texttt{hello} messages affecting the performance of an entire network. %Intrusion Detection Systems (IDS) utilizing Machine Learning (ML) or rule-based algorithms~\cite{andreoni2019toward} are employed for threat detection. 
Machine Learning (ML) based Intrusion Detection Systems (IDS)~\cite{andreoni2019toward} are employed for network threat detection. 
%MA: if necessary I can add more routing attacks such as Route flooding, Rushing attack, Route redirection, among others. 

%------------------------------------------------------------%
\subsection{Transport Layer}
\label{sec:threats_transport}

\boldpar{Vulnerabilities}
This layer controls data flow within a communication stack. Vulnerabilities in this layer are related to packet corruption or exploiting weaknesses within protocols. Transport Layer Protocol (TCP) (connection-oriented), User Datagram Protocol (UDP) (connection-less), and Internet Control Message Protocol (ICMP) are the main transport layer protocols.   

\boldpar{Threats}
In TCP every time a packet is sent with the SYN flag enabled, a new connection is opened. This is exploited on the \textit{TCP SYN Flood Attack} where an attacker creates a large number of TCP-SYN packets, creating a half-open connection to a target node spoofing the source address. On the victim side, every open connection exhausts memory, while the victim is not able to accept any other legitimate connection. 

In a \textit{UDP Flooding Attack} the attacker node sends a large number of UDP packets without payload to random ports on the victim, with a spoofed source IP address. On the victim side, when a service is not running on the port, a reply with an ICMP Destination Unreachable packet will be sent to the spoofed IP. Thus, the victim system sends many ICMP packets, being unreachable by other clients. 

In \textit{ICMP Flooding Attack}, also known as ping flooding, the attacker overwhelms the victim with ICMP echo requests, also known as pings. As this attack is very easy to reproduce using standard tools such as \texttt{hping3}, it is one of the reasons why corporations block ICMP packets in their network. 
% However, ICMP messaging is used within mesh protocols such as BATMAN and RPL

The \textit{Session Hijacking Attack} exploits the vulnerability of the transport protocols that do not check security during ongoing sessions. Thus, an attacker spoofs the IP address of a victim node, correctly determines the current sequence number that is expected to be generated at the victim node, and then performs a DoS attack on the victim node. Network threats are normally detected using IDS. Also, for protection software firewall systems can be used. Moreover, Transport Layer Security (TLS) can be used to encrypt the communication over this layer.

%------------------------------------------------------------%
\subsection{Application Layer}

\boldpar{Vulnerabilities}
The Robot Operating System (ROS) is a framework widely adopted for robotic middleware, as an example of the application layer in UAV Swarms. Since ROS was mainly designed for research purposes, no security features were added during its implementation.%~\myciteme.
The lack of security features such as data encryption and authentication make ROS a critical surface for different attack vectors in UAV networks~\cite{mcclean2013preliminary}. ROS2 is built on the top of Data Distribution Service (DDS) with a Real-Time Publish Subscribe (RTPS) architecture and was designed to provide a solution for the vulnerabilities found in its earlier iteration -- adding authentication, encryption, and process profile features, which rely on public key infrastructure. However, some vulnerabilities on the DDS system impacting the ROS2 key exchange were highlighted in~\citet{white2019network}. 

\boldpar{Threats}
Message Queuing Telemetry Transport (MQTT) is a lightweight, publish-subscribe protocol that transports messages between devices. Running on top of TCP/IP, it is typically used to send data from the drones to the cloud; its popularity is based on minimal bandwidth requirements and low memory consumption. Nevertheless, it has been shown that by default the MQTT protocol (i)~allows anybody to subscribe to the broadcasted topic without any authentication; (ii)~MQTT does not provide any data encryption and (iii)~The attacker who has already known the data packets by sniffing the traffic can modify the data in transit~\cite{anthraper2019security}. Moreover, Mosquitto (a popular MQTT broker) allows a malicious client to supply invalid data as shown in CVE-2017-7653
\footnote{https://cve.mitre.org/cgi-bin/cvename.cgi?name=CVE-2017-7653}. 
Additionally, \cite{sasi2015maldrone} shows how in an open application protocols such as Telnet and FTP, an attacker can connect, %through these protocols
execute commands, and transfer data~\cite{westerlund2019drone}. With these features, % If an attacker can send files and execute commands in the victim drone, 
a malware can be installed in the victim drone. A malware called Maldrone was created and tested in Parrot AR Drone 2.0 and a DJI Phantom. In the demo
% ~\footnote{demo of maldrone malware http://samy.pl/skyjack/}
the author showed that after malware installation, the drone can be forced to land in a specific location. 
DDS uses a naming convention for its topics including information about software or data type such as \textit{/my/image/raw\_image}.  Thus, an attacker can apply text recognition or natural language programming to fingerprint the device, un-patch or exploit versions of software/firmware, or even discover the network topology. 

\section{Security Architecture for UAV Mesh Networks %\textcolor{blue}{(Please read it)}
}
\label{sec:newarch}
\begin{figure*}
    \centering
    \includegraphics[width=1\textwidth]{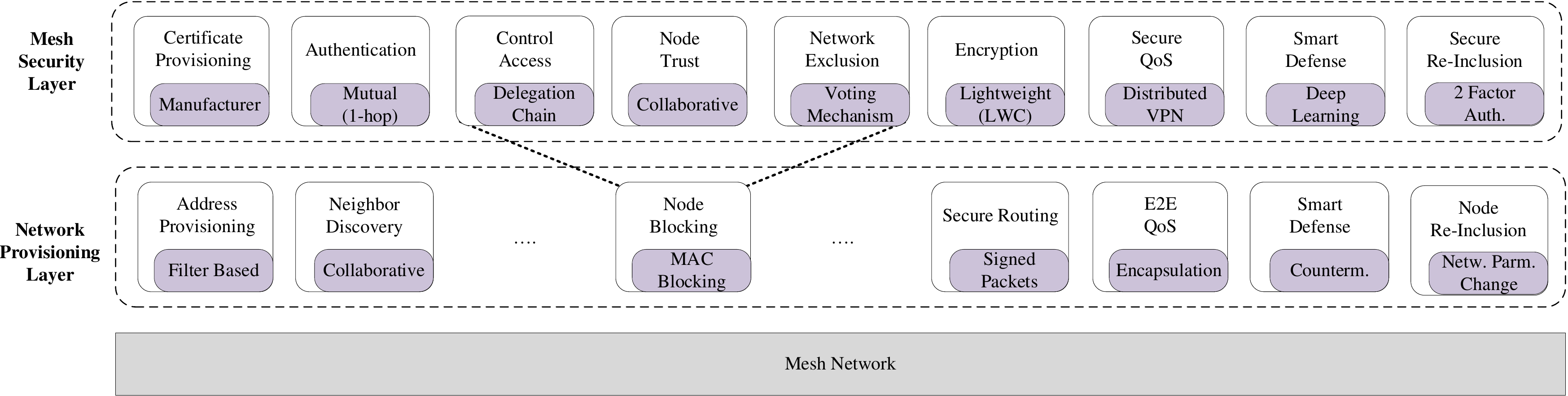}
    \caption{Secure architecture for future UAV swarm deployment. The architecture is composed of two layers: The Mesh Security Layer includes features for resilience and the Network Provisioning Layer is responsible for the implementation of the security features in the network stack. }
    \vspace{-4.00mm}
    \label{fig:new_architec}
\end{figure*}

The proposed security architecture addressing the issues highlighted in Section~\ref{sec:threats} is presented in Figure~\ref{fig:new_architec}. This architecture is designed to fill all the requirements of a fully distributed UAV mesh network. It is divided into two main layers: the \textit{Network Provisioning Layer} is responsible for managing the network aspects in a \eleven-based mesh network, while the \textit{Mesh Security Layer} includes features related to the detection and protection of the main threats for UAV swarms. 

\boldpar{Network Provisioning Layer}: \textit{Address Provisioning} is responsible to deliver a unique IP address to every participant node. %Assuming it is desirable to implement
In a fully distributed network where no central authority exists, address provisioning is a challenge. Filter-based techniques are proposed as a lightweight solution for mobile ad hoc networks in~\cite{fernandes2009efficient}. \textit{Neighbor Discovery} should be a collaborative, latency, and energy-aware solution to determine the participants and the way they are distributed on the network. \cite{wei2018fast}~propose a solution to combine these requirements in a group-based fast neighbor discovery algorithm (GBFA) by carrying neighbor information in a beacon packet.  

\textit{Node Blocking} must be a solution to avoid communication with any compromised node. Once the node is detected as compromised, a network countermeasure should be taken. 

On the other hand, a total exclusion from the network should be avoided, since it would be interesting to return the drone to the control base for forensic analysis. MAC address blocking is an effective solution using \texttt{ebtables}, however, it is susceptible to MAC spoofing attacks. \citet{ferraz2014accurate}~propose a method to exclude a malicious node both locally within the mesh and globally within the whole network. 

The \textit{secure routing} approach is a way to ensure that packets are not modified while passing the network, avoiding routing attacks. The simplest solution is to sign all the routing packets. This approach, however, can consume considerable resources within in a UAV swarm environment. \citet{sbeiti2015paser}~propose a combination of symmetric and PKI infrastructure for packet signing. 

The \textit{end-to-end Quality of Service (QoS)}, needs to be applied to ensure that main services in the mission have priority over the network without suffering any delay. 
Generic Routing Encapsulation (GRE) %~\cite{rfc2890}
% \footnote{Request for Comments 2890 https://tools.ietf.org/html/rfc2890} 
and Virtual eXtensible Local Area Network (VXLAN) %~\cite{rfc7348}
% \footnote{Request for Comments 7348 https://tools.ietf.org/html/rfc7348} 
are encapsulation methods in layers 2 and 3 used in traditional networks. Nevertheless, these approaches add an important overhead in the network traffic that can be harmful to UAV swarm networks. Thus, a lightweight solution must be used. \textit{Smart Defense} applies all countermeasures to avoid threats and intrusion to the network. 

\textit{Node Re-Inclusion} is a network method applied under situations where a node disappeared, returned, and must then be re-included in the network. Measures should be taken to verify if the node has been compromised. For example, all mesh parameters can be renewed to avoid data exfiltration. It is necessary to measure the impact of the parameter renewal when the drones are in the mission.

\boldpar{Mesh Security Layer}: \textit{Certificate Provisioning} must avoid communicating with the Certificate Authority during the mission. Initially, the certificate can be provisioned during the manufacturing process and key pairs are stored in Physical Unclonable Function (PUF), a hardware tamper-resistant approach. If the certificates are provisioned by the manufacturer, it is assumed the delegated trust route, trusting the entire web of trust until the producer root. \citet{blockchain}~propose a decentralized web-of-trust using blockchain. 

The \textit{Authentication} should use the certificates provisioned in the previous step. Furthermore, the authentication must be on both sides, each node should authenticate with its immediate neighbor mutually. \citet{zhu2003lhap}~propose a lightweight algorithm to authenticate all packets in hop-by-hop authentication. Moreover, \citet{jan2019payload}~suggest a mutual payload authentication method in a four-way handshake to avoid overhead in UAV swarm networks. Mutual Authentication is directly related to the neighbor discovery in the network layer since the nodes must know their neighbors to start the authentication process. The Control Access, Node Trusting, and Network exclusion are related to the node blocking scheme. %Once node A authenticates its neighbor, node B, node A will be responsible for the entrance of node B in the network. 
Once a particular node A authenticates a neighboring node B, it will be responsible for the entrance of node B in the network. If a node presents malicious or selfish behavior, 

\textit{Control Access} is mandatory for nodes' cooperation and the correct operation of the network. Therefore, a delegation chain of responsibility between node A and node B is created with access control. In the UAV swarm environment, the fog drone will be responsible for the authentication and control access of all nano-drones. 

\textit{Node Trust} is a mechanism to measure the misbehaving degree of the nodes to improve the network performance and needs to be collaborative between nodes. %, sharing information of possible malicious behavior to take a group decision. 

\textit{Network Exclusion} is the final step to remove a malicious node. In this step, a voting mechanism can be used to avoid malicious node collusion in the network. These three previous characteristics are addressed by~\cite{ferraz2014accurate}, which proposes an entire framework for control access based on node reputation. Encryption is necessary to protect the network from any kind of information leakage. Since UAV swarms can execute confidential missions, it is necessary to protect the packet payload but also the header information to avoid network and mission recognition. As mentioned before, traditional encryption methods can be energy intensive. Hence, Lightweight Cryptography (LWC) with smaller keys sizes and energy-awareness must be used in UAV swarm environments~\cite{lwc}. This feature is related to secure routing since the packet must be signed to fulfill the non-repudiation requirement. 

\textit{Secure Quality of Service} is directly related to the end-to-end QoS in the network layer. The objective of the Secure QoS is to provide end-to-end security to special services. To this end, tunnels can be created from the source to the destination. Dynamic tunnel creation is proposed in~\cite{juma2020hybrid} with a combination of Virtual Private Network (VPN) and the IPsec protocol for smart IoT objects. The dynamic tunnel creation is applicable in the UAV swarm environment, but an energy analysis must be performed. %In \textit{Smart Defense}, all the secure detection methods are applied to protect the UAV mesh network. Stream processing~\cite{andreoni2019toward} and Deep Learning approaches are used for network anomaly detection~\cite{hwang2020unsupervised}.

In \textit{Smart Defense}, all the countermeasures to protect the UAV mesh network after the detection of threats are applied. Stream processing~\cite{andreoni2019toward} and lightweight Deep Learning approaches are used for network anomaly detection~\cite{hwang2020unsupervised} and are combined with smart firewalls to protect the network. 

Finally, \textit{Secure Re-Inclusion} is the %security 
complement to re-incorporate a node into the network, avoiding spoofing or impersonation attacks. The PUFs can be used as the first factor of authentication and a pre-shared key as the second factor~\cite{gope2018lightweight}.

%\section{Challenges, Open Issues, and Future Research Directions}\label{sec:challenges} %\textcolor{blue}{(Michael, Willian, Anshul ...)} 
%\vspace{-6.00mm}
\section{Challenges and Opportunities}
\label{sec:challenges} %\textcolor{blue}{(Michael, Willian, Anshul ...)} 

% MB: Currently this table does no such thing
% Based on the opportunities outlined in Table~\ref{table:threats_and_vulnerabilities}, we provide an overview of the challenges and open issues in the application of wireless mesh networks within the field of UAV swarm communications. Specifically, 
% We provide an overview of challenges and open issues in the application of wireless mesh networks within the field of UAV swarm communications. 
While Section~\ref{sec:threats} demonstrates that many attacks and exploits can be mitigated with existing security mechanisms, we identify \textit{(i) high mobility}, \textit{(ii) vulnerability to jamming}, \textit{(iii) susceptibility to compromised nodes}, and (iv) \textit{positional awareness and security} as presenting immediate and critical vulnerabilities within UAV mesh networks %that are 
as key areas for future research.

% MB: Optimal routing for performant e2e applications, Synchronous Flooding for dependable network-wide communications and synchronization.
\boldpar{High UAV mobility} Distributed routing protocols such as OLSR~\cite{olsr} and BATMAN~\cite{ietf_batman} provide a measure of support for mobility through local decision making on neighbors and forwarding. While these solutions are typically employed within \eleven-based mesh networks and are theoretically able to handle the high throughput application traffic required in UAV mesh networks, such protocols were not designed for the extreme mobility experienced in (relatively recent) UAV use-cases; constantly changing topology and channel conditions can present significant challenges to mesh networks. While routing techniques can certainly be applied to static UAV formations, exploration of recent research into flooding techniques could allow reliable network-wide broadcast of mission and flight information within a swarm.
%Recent research in wireless mesh communications has resulted in a significant number of mesh routing protocols, with distributed routing protocols such as OLSR~\cite{olsr} and BATMAN~\cite{ietf_batman} providing a measure of support for mobility through local decision making on neighbors and forwarding. While these solutions are typically employed within \eleven-based mesh networks and are theoretically able to handle the high throughput application traffic required in UAV mesh networks, such protocols were not designed for the extreme mobility experienced in (relatively recent) UAV use-cases; constantly changing topology and channel conditions can present significant challenges to mesh networks. While routing techniques can certainly be applied to static UAV formations, exploration of recent research into flooding techniques could allow reliable network-wide broadcast of mission and flight information within a swarm.

Within the low-power wireless community Synchronous Flooding (SF) protocols have been used to \textit{intentionally} let multiple relaying nodes forward packets without fear of collision by synchronously and simultaneously broadcasting them on the same carrier frequency~\cite{zimmerling20synchronous}. This enables protocols based on SF to reliably flood a mesh with \textit{the minimum possible latency dictated by the physical layer}. Through exploitation of non-destructive interference and the capture effect, this flooding technique has consistently been shown to outperform routing solutions in terms of reliability and latency in low-power mesh networks~\cite{schuss17competition}. While SF is limited to frequency demodulated physical layers~\cite{baddeley2020impact} and thus cannot support high-throughput applications such as video, the promise of interference mitigation, low-latency, network-wide synchronization, and agnosticism to mobility, make it a promising candidate for time-critical control communications within a UAV swarm. 

% CTs rely on two physical layer phenomenon present in frequency-modulated systems to work: non-destructive interference the capture effect~\cite{leentvaar76capture}. The former allows CT-based transmitters to simultaneously send the \textit{same data} to support reliable broadcasting for over-the-air updates, whilst the latter allows correct reception of simultaneously transmitted \textit{different data}, e.g., to support data collection services towards a border router.

% MB: FHSS, multi-radio communications, intelligent anti-jamming techniques (spectrum sensing)
\boldpar{Vulnerability to RF jamming}
The growing hardware power and falling cost of Software Defined Radio (SDR) kits alongside the widespread availability of ML techniques make sophisticated \textit{reactive} and \textit{cognitive} jamming attacks far more accessible. However, there still exist limitations in the capability of many RF jamming approaches. \textit{Reactive} jammers will take a few milliseconds to ramp up their radio in response to a transmission. As such, a small amount of data will be demodulated at the receiving device before the reactive jamming signal kicks in.  For example, 1 byte of data transmitted over a 1\,Mbps physical layer (PHY) has an on-air time of 8\,$\mu$s. Non-standard PHY and MAC implementations could therefore avoid such jamming techniques by keeping packets short in time, combined with standard diversity techniques such as frequency, space, time. Furthermore, while the addition of FHSS-based MAC options to standards such as \fifteenfour~\cite{802154_ieee} have resulted in many protocols that can escape external interference through channel hopping across a spectrum band, this may still be insufficient to defend against deliberate jamming attacks. However, multi-radio approaches could allow swarm communications across multiple bands and therefore mitigate this threat. 

% MB: Continuous authentication and PHY-layer security both come under susceptibility to compromised nodes
\boldpar{Susceptibility to compromised nodes} 
The highly dynamic environments associated with UAV use-cases lend weight to a need for flexible and decentralized security infrastructure. Traditional authentication schemes authenticate devices upon joining a network: deciding whether it is authenticated or not. Therefore, they are vulnerable to security threats such as hijacking, which take control of active sessions. To address this, Continuous Authentication (CoA) has been seen as a solution to maintaining authentication of UAV node identity~\cite{shoufan2017continuous}. However, as CoA is performed frequently and the IoT devices are resource-constrained, it is necessary to adopt a lightweight mechanism to reduce their energy consumption~\cite{yahuza2021edge}. This further complicates with the consideration of multi-hop traffic within a wireless mesh network.

Physical layer security (PHY-security) techniques are seen as a possible solution to achieve this. These techniques exploit the inherent random characteristics of the wireless channel (such as multipath fading, interference, propagation delay, etc.,) to realize key-less secure transmissions through various coding and signal design and processing techniques~\cite{phy_urllc}, and ML-based approaches~\cite{yu2019radio}.  These solutions guarantee information security by  degrading the channel capacity of the adversary, and does not need any third party intervention. In the context of UAV mesh communications, PHY-security solutions can provide mutual authentication~\cite{phy_auth}, lightweight encryption~\cite{phy_key}, and anti-eavesdropping strategies~\cite{adaptive_phy,phy_tech}. Specifically, physical layer-assisted CoA schemes are designed by exploiting the unique features of a specific transceiver pair (channel state information, received signal strength indicator (RSSI)) and/or analog front-end imperfections. Moreover, the inherent randomness of the wireless channels and the reciprocity principle can be used for physical layer-assisted key generation to provide continuous point-to-point encryption. 
Furthermore, techniques like beamforming and precoding~\cite{phy_beam}, multi-user and multi-relay diversity~\cite{phy_mur}, relay cooperation~\cite{adaptive_phy}, and artificial noise generation~\cite{phy_tech} can degrade the channel quality of the malicious users and thus reduce information leakage. Thus, PHY-security is a potential solution to strengthen the security infrastructure in UAV wireless mesh networks. 

\boldpar{Positional awareness and security} 
Global Navigation Satellite System (GNSS) services have traditionally provided positioning and synchronization for UAVs, both of which are critical for mission and flight control. However, such services are designed for outdoor scenarios with direct satellite links, therefore precluding them from indoor environments. Furthermore, with low SNR, %they've been shown to be 
they are particularly vulnerable to malicious jamming and/or spoofing~\cite{gao2016protecting,psiaki2016gnss}. To address these issues, several works have examined Computer Vision-based odometry~\cite{caballero2009vision} to augment traditional GNSS. Recently, several works have shown how, by using relatively low-power Ultra Wideband (UWB) devices, one can achieve centimeter accurate localization and outperform BLE as well as \wifi devices. Such accuracy has been demonstrated as sufficient to efficiently navigate quadrotors and robots in indoor settings. However, it has been shown that the single or double-sided two-way ranging (TWR) schemes typically employed in UWB localization do not scale a key challenge in the context of UAV swarms. Inspired by the SF techniques~\cite{zimmerling20synchronous} previously mentioned in this section, many researchers are investigating how to exploit UWB-based concurrent transmissions for secure and scalable localization services. Specifically, it has been demonstrated that it is feasible to exploit this technique together with channel impulse response information for ranging, allowing passive self-localization to support countless targets~\cite{grobetawindhager2019snaploc}.
%\section{Discussion and Future Work} 
\section{Conclusion and Future Work} %MA: I did it more in the sense of conclusion. We can discuss it. 
%\textcolor{red}{(Leave until end.)}}
\label{sec:conclusions}
% Such a framework of 
UAV swarms employ wireless mesh to provide ad hoc networking infrastructure, both in simple applications such as geographical mapping or complex events like climate disasters or military missions. 
% The use of communication between drones, 
However,  weaknesses inherent within mesh protocols and standards, %introduces 
have introduced new attack surfaces and presents novel security challenges. In this paper, we have therefore highlighted the main vulnerabilities and threats across in the entire UAV mesh communication stack -- from the physical to the application layer. 
% In addition, we have discussed the fundamental challenges and open issues in UAV swarm in terms of security, resilience, and network optimization, and its importance for future research. 
In addition, we have discussed the fundamental challenges and open issues currently faced by UAV swarm communications and provide a road map for future research opportunities within this area.
% We have proposed a secure mesh communication architecture with all the mentioned requirements for resilience and network improvement. The architecture is composed of a network provisioning layer and a mesh security layer and details all the solutions for an improved UAV swarm secure communication. 
From this analysis, we have proposed a secure mesh communication architecture, consisting of the specified requirements for resilience and network improvement. It is composed of both a network provisioning and mesh security layer, and details solutions for improving security at all layers within UAV swarm communication stack.
% As future work, we intend to implement the proposed architecture and to execute the items presented as future directions. 
%As a future work, we intend to implement the proposed architecture and address the open challenges discussed within this paper.
As a future work, we will implement the proposed architecture and address the open challenges discussed within this paper.
% Moreover, another future work is to extend the proposed architecture to coordinate several types of UAVs, from fog to nano, that present different resource constraints.
%Moreover, this work will also extend the proposed architecture to address several categories of UAVs: presenting different resource constraints and use-case scenarios.
Moreover, this work will also extend the proposed architecture to address several categories of UAVs with  different resource constraints and use-case scenarios.
\vspace{-1.50mm}
\bibliographystyle{IEEEtranN}
\bibliography{biblio}

% that's all folks
\end{document}